\documentclass[aps,twocolumn,groupedaddress,prb]{revtex4}

\begin{document}
\bibliographystyle{ajp}


\title{Modernizing Newton, to work at any speed}


\author{P. Fraundorf}
\affiliation{Physics and Astronomy/Center for Molecular Electronics, U. Missouri-St. Louis, MO 63121}


\date{\today}

\begin{abstract}
Modification of three ideas underlying Newton's original world view, with only minor changes in context, might offer two advantages to introductory physics students.  First, the students will experience less cognitive dissonance when they encounter relativistic effects.  Secondly, the map-based Newtonian tools that they spend so much time learning about can be extended to high speeds, non-inertial frames, and even (locally, of course) to curved-spacetime. 
\end{abstract}

\maketitle

\tableofcontents

\section{Introduction}
\label{sec:Intro}
Relativistic environment dwellers in the earth's gravitational well often experience time differently than space.  Practical challenges for them thus begin and end in frame-specific terms\cite{Bell87}, even if coordinate-free insights help visualize the symmetry of the problem\cite{Misner73}.  Hence problem solvers in such environments can benefit from Newton-like laws [with units] based in their own frame of reference, that model nature over as wide a range of (location and velocity) conditions as possible\cite{Ashby94}.  

Just as some fluid-flow tools that we offer to introductory students don't apply under supersonic conditions, so the tools for thinking about motion fail in fundamental ways at relativistic speeds and in curved spacetime.  We discuss here ways to hedge our bets with the tools used to describe motion -- looking at the pros and cons of some changes in emphasis that can minimize the pain of deeper understanding later.  These changes might also open the door to wider application of frame-based tools in projects involving high speeds and/or extreme spacetime curvature.  For example, few children know that relativity enhances their options for long range travel in one lifetime, instead of limiting their travel to places within 100 light years of their place of birth.  Likewise, describing the world land speed record for particle acceleration in terms of ``map distance per unit traveler time'' could augment public interest in accelerators if particle speeds are expressed in units of lightyears per traveler year, the current land speed record (e.g. for 50 GeV electrons) being on the order of $10^5$!

Access to improved tools can be facilitated by avoiding three assumptions implicit in the classical worldview:  (i) global time, (ii) frame-invariant dynamics, and (iii) limitation to inertial frames.  These assumptions generate cognitive dissonance when one later considers either high speed motion or modern views of gravity.  Their elimination does not require major changes.  It further facilitates the extension of familiar Newtonian tools to problems involving acceleration at high speeds, and (with help from the equivalence principle) to arbitrary coordinate systems as well.

\section{Misconception One}

``Time is the same for everyone.''  This is the implicit assumption of global time, against which most spoken languages have little built in protection. 

\subsection{The Specified-Clock Fix}

Make a habit of specifying the clock used to measure time $t$, as well as the coordinate system yardsticks used to measure position $x$.  With time specified as measured on map-clocks (i.e. a set of synchronized clocks connected to the yardsticks with respect to which distance is measured), the usual kinematic equations of introductory physics need not be changed, even if their precise meaning (at least in the case of equations describing relative motion) is different.

\subsection{Consequences of Specified-Clocks}

When $t$ represents time on ``map-clocks'', the usual definitions of velocity and acceleration (as derivatives of map-position with respect to $t$) define what relativists sometimes call ``coordinate'' velocity and acceleration, respectively\cite{Misner73}.  These definitions are useful at any speed, although they become less global when spacetime is curved.  If the integrals of constant acceleration, like $\Delta v = a \Delta t$ and $\Delta (v^2) = 2 a \Delta x$, represent relations between map-time and map-position, they follow simply from these definitions and hence are also good at any speed.  The caveat, of course, is that holding coordinate acceleration finite and constant is physically awkward at high speeds, and impossible to do forever.  This diminishes the usefulness of these equations in the high speed case\cite{French68}, but not their correctness.  Thus being specific about clocks yields classical definitions of velocity and acceleration, plus familiar equations of constant coordinate acceleration, that work even at relativistic speeds.

The kinematical equations for {\em relative motion} (e.g. $x' \cong x + v_o t$ and its time derivatives) now refer only to yardsticks and clocks in a single reference frame.  Hence they only tell us about ``map-frame'' observations.  While they are correct in these terms, the prediction of prime-frame perspectives is only valid at low speed.  Since multiple inertial frames are a challenge in and of themselves\cite{Scherr01}, simply 
describing it as a low-speed approximation may be the best choice in an introductory course.  Nonetheless much can be done from the perspective of a single inertial frame to give quantitative insight into motion at high speeds.  Exact position, velocity, and acceleration transforms from the metric equation (like relativistic velocity addition and the Lorentz transform) thus may not provide the most useful insights in limited time.  In any case with subtle changes like this, introductory physics courses can become natural venues for amending the assumption of ``global-time'' implicit in the wiring of our brains.  

\subsubsection{Extension: Fast Airtracks}

Rather than simply ``being told'' the metric equation and ways to use it, students might benefit from time spent first pondering how to explain data on traveler/map time differences, in the spirit of Piaget and modeling workshop\cite{Hestenes92}.  In that case an airtrack thought experiment, with adjustable kinetic-energy source (e.g. a spring) and two gliders that stick on collision, is adequate to give students data for experimental discovery (in their own terms) of phenomena underlying the metric equation as well as relativistic expressions for the kinetic energy, $K = (\gamma - 1) m c^2$, and momentum, $p = m w$, of a traveler of mass $m$. \cite{Fraundorf01a}  A particular student's exploration might for example uncover, instead of the metric equation, the mathematically equivalent relation that holds constant the sum of squares of their ``coordinate speeds'' through space ($dx/dt$) and time ($c d \tau /dt$).

\subsubsection{Extension: One-Map Two-Clocks}

The conceptual distinction between stationary and moving clocks opens the door for students to discover, and apply,  Minkowski's spacetime version of Pythagoras' theorem\cite{Minkowski08} $(c \Delta \tau )^2 = (c \Delta t )^2 - (\Delta x)^2$, i.e. the metric equation, as an equation for time $\tau$ on traveler clocks in context of coordinates referenced to a single map frame\cite{Taylor98}.  Lightspeed $c$ here acts simply as a spacetime constant that connects traditional units for distance and time.  The metric equation connects (i) traveler and map time intervals, (ii) coordinate velocity, $ v \equiv dx/dt$, with proper velocity\cite{SearsBrehme68}, $ w \equiv dx/d \tau$, and (iii) special to general relativity as well.   

One consequence of such an introduction to clock behavior at high speed, for example, is that an upper limit on coordinate-velocity $v$ may appear more natural to students since it follows from the lack of any such limit on proper-velocity $w$.  One would hardly expect this latter quantity (map distance traveled per unit traveler time) to exceed infinity for a real world traveler.  The absence of an upper bound also makes a proper-velocity of one [lightyear per traveler year] at $v=c/\sqrt{2}$ a natural ``scale speed'' for the transition from sub-relativistic to relativistic regimes.  The distinction between traveler and map time also yields a natural definition for Einstein's gamma factor as a speed of map-time per unit traveler-time given simultaneity defined by the map, i.e. $\gamma \equiv dt/d \tau$.  The metric equation then easily yields the familiar relations, $\gamma = 1/ \sqrt{1 - (v/c)^2} = w/v$.  

\subsubsection{Drawbacks: Irrelativity}

The downside of specifying clocks, of course, is that global time is more deeply-rooted and simpler than clock-specific time.  For example, we often treat magnetism and gravity as non-relativistic add-ons to a classical world without giving it a second thought.  On the other hand, our present understanding of these phenomena grows from a metric equation that looks like Pythagoras' theorem with a minus sign.  Thus {\em only} tradition tells us that relativistic effects don't color our lives everyday.

\section{Misconception Two}

``Coordinate acceleration and force are frame-invariant.''  Informal polls at one state\cite{Fraundorf97c} and one national\cite{Fraundorf98a} AAPT meeting suggest that physics teachers often presume that the rate of change of conserved quantities (energy as well as momentum) is frame-invariant, even though the expression for power as a product of force and velocity makes this clearly incorrect at even the lowest speed.  

\subsection{The Subjective-Dynamics Fix} 

Address the issue of frame invariance when dynamical quantities are first introduced.  Point out that the value of the conserved dynamical quantities momentum and energy, along with their time derivatives force and power, depends on the frame of reference (even though force is nearly frame-independent at low speeds).  Likewise, note that coordinate acceleration $a = dv/dt$ is not necessarily the acceleration {\em felt in the proper frame of a traveler}.  The former is impossible to hold finite and constant indefinitely, while proper acceleration $\alpha$ is both frame-invariant and possible to hold constant\cite{Lagoute95}.  This is what Newton had guessed (incorrectly) to be the case for coordinate acceleration, by a perfectly reasonable application of Occam's razor given the observational data that he had to work with at the time\cite{Newton46,Chandrasekhar95}.

Quantifying these accelerations at arbitrary speed is simple in the unidirectional case\cite{Mallinckrodt99}, where $\alpha = \gamma ^3 a$.  An invariant proper-force $F_o = m \alpha$ may also be defined in this context\cite{Taylor63}.  Quantifying the conserved dynamical quantities further requires relativistic expressions for momentum $p = m w$ and energy $E = \gamma m c^2$.  Their derivatives with respect to map-time then become the familiar net frame-variant force $\Sigma F = m dw/dt$ and power $dE/dt = \Sigma F \bullet v$.  

These quantities most directly represent the way that momentum and energy associated with a traveling object is changing, from the perspective of a map-frame observer.  Proper-time derivatives of momentum and energy are of course easier to transform from frame to frame (they form a 4-vector), but they also represent a perspective intermediate between that of the map-frame observer and the traveling object.  The focus here is on local implementation of the insights in terms of map-frame observer and traveling-object experiences.  Hence references here to coordinate-free insights that underlie the conclusions (e.g. to the four-vector nature of certain quantities) are for instructor reference, but of limited use to students wanting a sense of anyspeed motion only in concrete terms.

Perhaps coincidentally in the unidirectional motion case, the net frame-variant force $\Sigma F$ and the proper-force $F_o$ are equal.  In flat (3+1D) space-time, proper acceleration and proper force behave like a 3-component scalar (i.e. no time component) within the ``local space-time coordinate system'' of an accelerated object\cite{Misner73}, while the corresponding frame-variant forces differ and may be accompanied by frame-variant energy changes.  The potentials that give rise to such frame-variant forces inevitably have both time-like (e.g. electrostatic) and space-like (e.g. magnetic) components\cite{Minkowski08}.  In curved space-time and/or rotating coordinate systems, the proper-acceleration and proper-force retains it's three-component form only from instant to instant, within the traveling object's ``proper reference frame''\cite{Misner73}.

\subsection{Consequences of Subjective-Dynamics}

To begin with, students are alerted to the fact that relativity makes many quantities 
more dependent on one's choice of reference frame.  For some of these quantities (time 
increments, object length, mass\cite{Adler87}, traveler velocity\cite{Shurcliff96}, and acceleration\cite{Taylor63}) there is a ``minimally variant'' or proper form which can simplify discussion across frames.  

\subsubsection{Extension: Proper Acceleration}

Recognizing the distinction between coordinate and proper acceleration, students also 
gain (from the metric equation) simple 1D equations for describing constant 
proper-acceleration.  In terms of coordinate integrals like those for constant 
coordinate-acceleration above, there are three equations instead of two\cite{Fraundorf97b} because the metric equation relates three coordinates:  map-time $t$, map-position $x$, and traveler-time $\tau$.  The integrals are $\Delta w = \alpha \Delta t$, $c^2 \Delta \gamma = \alpha \Delta x$, and $c \Delta \eta = \alpha \Delta \tau $ where the ``hyperbolic velocity angle''\cite{Taylor63}  or rapidity $\eta \equiv \tanh^{-1}[v/c]$.  One can also give students a set of map-based Newton-like laws good in (3+1)D flat spacetime, with only a modest amount of added complication\cite{Fraundorf97a}.  

For example, Newton's second law for flat spacetime looks like
\begin{equation}
{\mathord{\buildrel{\lower3pt\hbox{$\scriptscriptstyle\leftarrow$}} 
\over F} } = m \frac{\alpha }{{\gamma _t }}\mathord{\buildrel{\lower3pt\hbox{$\scriptscriptstyle\leftarrow$}} 
\over i} _l \mathop = \limits^{v_{\perp} \ll c} m \mathord{\buildrel{\lower3pt\hbox{$\scriptscriptstyle\leftarrow$}} 
\over \alpha } \mathop = \limits^{v \ll c} m \mathord{\buildrel{\lower3pt\hbox{$\scriptscriptstyle\leftarrow$}} 
\over a} 
\end{equation}
where column three-vectors are denoted by left-pointing arrows, and unit vectors are denoted by the letter $i$.  As usual $v$ and $a$ denote 
coordinate-velocity and coordinate-acceleration, while $\alpha$ denotes proper-acceleration.  Vector components transverse and longitudinal to the frame-variant force direction are denoted by subscripts $t$ and $l$, respectively, while components perpendicular and parallel to the frame-invariant proper-acceleration three-vector are denoted by subscripts $\perp$ and $\parallel$.  Also $\gamma_t \equiv \sqrt{1+(w_t/c)^2}$, and the unit vector longitudinal to the frame-variant force is related to unit vectors perpendicular and parallel to the proper-acceleration by
\begin{equation}
\mathord{\buildrel{\lower3pt\hbox{$\scriptscriptstyle\leftarrow$}} 
\over i} _l  = \gamma _t \gamma _ \bot  \left( {\frac{1}{{\gamma _ \bot  ^2 }}\mathord{\buildrel{\lower3pt\hbox{$\scriptscriptstyle\leftarrow$}} 
\over i} _{\parallel}  + \frac{{v_ \bot  }}{c}\frac{{v_{\parallel} }}{c}\mathord{\buildrel{\lower3pt\hbox{$\scriptscriptstyle\leftarrow$}} 
\over i} _ \bot  } \right),
\end{equation}
where $\gamma_{\perp} \equiv 1/\sqrt{1-(v_{\perp}/c)^2}$. 
Hence the frame-variant force direction differs from that 
of the invariant proper acceleration only when the velocity of 
the object being effected is neither perpendicular to, nor 
parallel to, the proper acceleration.

\subsubsection{Extension: Galileo's Chase-Plane}

Relativistic equations for constant proper acceleration work 
poorly at low speeds because of roundoff error in calculating 
the difference between squares, while Galileo's equations \cite{Galileo62} 
(which students spend so much time learning) are elegant in 
their simplicity.  Perhaps it is good news then that by 
going from ``one-map two-clock'' descriptions of motion {\em ala} the 
metric equation to one-map and three-clocks, Galileo's 
equations can be shown to apply to constant proper-acceleration 
at any speed.  

Specifically, in terms of time T on the clocks 
of a suitably-motivated ``chase-plane'', one can show \cite{Fraundorf96a} that 
the Galilean-kinematic velocity $V \equiv dx/dT$ of a traveler 
undergoing constant proper-acceleration $\alpha$ obeys 
$\Delta(V^2) = 2 \alpha \Delta x$, $\Delta V = \alpha \Delta T$, 
etc.  This familiar and simple time evolution seamlessly 
bridges the gap to low speeds since $v < V < w$.  It also predicts the 
experience of relativistic observers using the classic equations 
since, for example, at any speed  
$\gamma = 1 + \frac{1}{2} (V/c)^2$ and proper velocity 
$w = V \sqrt{1 + \frac{1}{4} (V/c)^2}$.

\subsubsection{Extension: One-Frame Magnetism}

The non-parallel relationship between frame-variant force 
and proper acceleration is manifest in our everyday life 
as a need to recognize both electrostatic and magnetic 
components to the Coulomb force between moving electric charges.  
This for example makes possible single-frame (one-map 
two-clock) derivations (cf. Appendix \ref{AppxA})
of the Lorentz Law and Biot-Savart in (3+1)D.

\subsubsection{Drawbacks: Inter-Scale Tension}

One downside of discussing the fact that observed forces depend on one's frame of motion (in flat spacetime) and one's location (in curved spacetime) is the elegant simplicity (and for most engineers, the elegant practicality as well) of ``global'' force and acceleration in the Newtonian worldview.  Refinement of prose to break this news to students in context (the ``detail work'' of content modernization) may therefore require a long period of experimentation and refinement.  Even then, true global perspectives on frame invariance (and its absence) will likely remain a corollary rather than a pillar of introductory dynamics.

\section{Misconception Three} 
``Newton's laws work only in unaccelerated frames.''  This misconception is 
often echoed in classes whose first example of a force is the affine-connection 
force\cite{Misner73} gravity, which like centrifugal force arises only if one chooses a 
``locally non-inertial'' coordinate system.  Of course this would be fine 
for historical reasons, if the equivalence principle hadn't elegantly shown 
that Newton's laws are useful {\em locally} in any frame\cite{Einstein50}.  This is potentially motivational 
news for those first struggling to understand how Newton's laws work, and 
perhaps worth sharing from the start.  

\subsection{The Equivalence Fix} 

Show examples of the way that Newton-like laws work locally in any frame, when one includes ``geometric'' (affine-connection) forces (e.g. centrifugal or gravity) that act on every ounce of one's being.  Geometric forces also have the property that they may be made to vanish at any point in space and time, by choosing a ``locally inertial'' coordinate system.  Non-local effects, like tides and coriolis forces, of course may not be possible to eliminate\cite{Misner73}.

Example 1:  On traveling around a curve of radius $r$ in a car at speed $v$, note that a weight suspended by a string from the rear view mirror accelerates away from the center of the turn.  In the non-inertial frame of the moving car, this is instinctively seen as the consequence of a ``centrifugal force''.  A more careful look shows that this force seems to act on every ounce of the object's being.  For example, the resulting acceleration is to first order independent of object mass (equal to $v^2 /r$), and it does not push or pull just on one side or the other.  Secondly, the explanation is only useful locally.  If an object is allowed to travel too far under this geometric force (e.g. more than 30 cm when going around a 30 m radius curve) complications arise in its motion not expected from a simple radially-outward force.  Finally, note that this force vanishes if one observes events from the ``locally-inertial'' frame of a pedestrian standing by the side of the road.  Free objects in the car are simply trying to move in a straight line in the absence of any force at all.

Example 2:  When standing near the surface of the earth, note that when you drop an object, it falls.  This is instinctively seen as a ``gravity force''.  On closer inspection, this force acts on every ounce of an object's being in that it gives rise to an acceleration ($g$) that is independent of mass.  No single part of the object seems to be pulled or pushed preferentially.  Note also that this force vanishes if one observes events from the ``locally-inertial'' frame of a person falling with the object.  To it's falling companion, the object is simply trying to move in a straight line, in the absence of any force at all.

Denying the usefulness or reality of either of these forces denies the utility of the equivalence principle itself.  The question is not if these forces exist.  One's sense of being forced to the side of vehicles as they round curves is as real as one's sense of being forced to the ground when a chair leg breaks.  Rather, we might better be asking:  What is the range of positions and times over which such ``geometric forces'' can be seen to govern motion in their non-inertial setting, while remaining consistent with a set of Newton-like rules.

\subsection{Consequences of Equivalence} 

Students are readied for simple equations that 
compare centrifugal and centripetal perspectives on travel around curves\cite{Misner73}, and electrostatic and gravitational point sources with regard to their relativistic (e.g. magnetic and curvature) effects\cite{Fraundorf97a}.  They also gain ``geodesic frame'' (e.g. satellite) and ``shell-frame'' (e.g. earth) based Newton-like equations of motion at any speed, around objects of any mass.  This is relevant to global positioning system applications which are forced to recognize the subjectivity of earth-based NIST clocks\cite{Ashby98}, and extreme environments like those discussed in Taylor and Wheeler's most recent text\cite{Taylor01}.

\subsubsection{Extension: Anyspeed Carousels}

Consider a set of fiducial (map-frame) observers who find themselves 
rotating at angular velocity $\Omega$ along with a set of yardsticks 
arrayed around the circumference of a circle of radius $r$.  In this 
case, the metric tensor 
for $x^0 = c t = c t_o \sqrt{1-(\frac{\Omega r}{c})^2}$, $x^1 = r = r_o$, 
$x^2 = \phi = \phi_o + \Omega t$, and 
$x^3 = z = z_o$ becomes 
\begin{equation}
g_{\mu \nu} = \left( {\begin{array}{*{20}c}
   { -\left[ {1 - \left( {\frac{{\Omega r}}{c}} \right)^2 } \right]} & 0 & {\frac{\Omega r}{c}}r & 0  \\
   0 & 1 & 0 & 0  \\
   {\frac{\Omega r}{c}}r & 0 & r^2 & 0  \\
   0 & 0 & 0 & 1  \\
\end{array}} \right) .
\end{equation}
As shown by Cook \cite{Cook04}, the spatial metric defining
local radar distance then assumes the decidedly non-Euclidean form
\begin{equation}
\left( {d\ell } \right)^2  = \left( {dr} \right)^2  + \frac{1}{{1 - \left( {\frac{{\Omega r}}{c}} \right)^2 }}\left( {r d\phi } \right)^2 + \left( {d z}\right)^2
\end{equation}
Thus distance is 
the same in the radial direction as in a 
stationary frame, while circumferential 
yardsticks are length-contracted in 
the azimuthal direction, increasing local 
radar distance as appropriate for
an azimuthal velocity of $\Omega r$ and 
illustrating contraction-effects within 
one (azimuthal ring) frame.

To find the geometric forces, we calculate 
the affine-connection and resulting geodesic equation 
in terms of coordinates in this rotating frame.  
Thus free objects experience a radial acceleration of 
the form 
\begin{equation}
\frac{{d^2 r}}{{d\tau ^2 }} = r\left( \gamma \Omega - \frac{d \phi}{d \tau} \right)^2, 
\end{equation}
where $\gamma$ is as usual $dt/d\tau$.  As any car passenger 
can attest, accelerations like this are experienced as forces 
that act on every ounce of one's being.  For observers at fixed $\phi$, 
this ``affine-connection force'' is a relativistic version of the 
familiar centrifugal force felt as one goes around curves in a car.  
This treatment of the problem goes further even for low-speed 
application.  For observers with fixed $r$ and changing $\phi$ e.g. 
for azimuthal track runners in a space station with artificial 
gravity, the above equation predicts a change in their 
``centrifugal weight'' depending on how fast and in which direction 
they run.  In particular by running quickly enough in a direction 
opposite to the satellite's rotation, they can make themselves 
weightless.

\subsubsection{Extension: Extreme Gravity}

The metric tensor for $x^0 = c t$, $x^1 = r$, $x^2 = \theta$, 
and $x^3 = \phi$, here in ``far coordinates'', becomes
\begin{equation}
g_{\mu \nu} = \left( {\begin{array}{*{20}c}
   { - (1 - \frac{2 G M}{c^2 r})} & 0 & 0 & 0  \\
   0 & \frac{1}{1-\frac{2 G M}{c^2 r}} & 0 & 0  \\
   0 & 0 & r^2 & 0  \\
   0 & 0 & 0 & r^2 \sin[\theta]^2  \\
\end{array}} \right) .
\end{equation}  
Although this space-time curvature only changes the 
metric coefficients by about a part per billion at 
the surface of the earth, it beautifully explains 
much of what we experience about gravity today, and more.

Of course, the equation only applies exterior 
to a spherically symmetric mass M.  For objects whose 
mass lies within the {\em event horizon} radius predicted by 
this metric, at $r = \frac{2 G M}{c^2}$, the metric 
also only applies exterior to the event horizon as 
well.  Cook \cite{Cook04} shows (strangely enough) that 
for both ``shell frame'' (r constant) and ``rain frame'' 
(free falling from infinity) observers, the local physical 
metric (i.e. Pythagoras' theorem) remains Euclidean, at 
least outside the event horizon.  This in spite of the 
fact that the two are obviously in different states of 
acceleration.  

Calculating far-coordinate affine-connection terms, 
and the resulting geodesic equation, predicts a radial 
acceleration for stationary objects of the form
\begin{equation}
\frac{{d^2 r}}{{d\tau ^2 }} = \frac{G M}{r^3} \left( r - \frac{2 G M}{c^2} \right) . 
\end{equation}
This acceleration in far-coordinates thus for example 
goes away (because of the apparent slowing down of time) at the 
event horizon, but cleanly reduces to Newton's gravity law for 
$r \gg  2 G M/c^2$.

\subsubsection{Drawbacks: Life on a Shell}

Disadvantages of discussing the local validity of Newton's laws, as distinct from their correctness only in unaccelerated frames, are at least twofold.  To begin with, the concept of ``local validity'' is rather sophisticated.  Admittedly we have many concrete examples, like the local validity of the uniform gravitational field approximation at the earth's surface, and it's inability to deal with less local phenomena such as orbits and lunar tides.  A deeper problem is the difficulty of explaining what an inertial (e.g. rain) frame is, if stationary frames (e.g. sitting down on a stationary earth) can be non-inertial.  It is simpler (or at least more traditional) to introduce inertial frames by referring to their uniform motion relative to some inertial standard, rather than by referring to the absence of a felt ``geometric'' acceleration acting on every ounce of mass in one's corner of the world.

\section{Conclusions}

Measurement of time and mass in meters provides relativistic insight into global symmetries, and the relation between 4-vectors in coordinate-independent form.  However, frame-specific Newton-like laws with separate units for length, time and mass are perhaps still crucial to the inhabitants of any particular world, as an interface to local physical processes.  

We point out some advantages of the fact that Newton's laws, written in context of a map-frame of choice, have considerable potential beyond their classical applications.  By avoiding the implicit assumptions of (i) global time, (ii) frame-invariant dynamics, and (iii) limitation to inertial frames, introductory students can be better prepared for an intuitive understanding of relativistic environments, as well as for getting the most out of the laws themselves.  

\begin{acknowledgments}
Several decades of energetic work by E. F. Taylor, on pedagogically-informed content-modernization, has offered key inspiration.
\end{acknowledgments}

\appendix

\section{One-Frame Biot-Savart}
\label{AppxA}

By way of application, imagine a wire running from left to right with a positive charge density of $+\frac{e}{\ell_o}$ and a negative charge density of $-\frac{e}{\ell_o}$, where $\ell_o$ is the distance between charges of given sign.  On a chunk of wire of length $ds$, this translates to a positive charge $Q = \frac{e}{\ell_o} ds$ and a negative charge of $-\frac{e}{\ell_o} ds$ and therefore a net charge of $0$.  If the positive charges are stationary in the wire, but the negative charges are moving to the right with a speed $v$, we say that the current in the wire is $I = \frac{e}{\ell_o} v$, to the left.  From Coulomb's law, the electrostatic force on a stationary test charge $q$ a distance $r$ above the wire is of course
\begin{equation}
F_{up} = F_{+}+F_{-} = k \frac{q Q}{r^2} - k\frac{q Q}{r^2} = 0
\end{equation}

Since the test-charge is not moving, the flat-space version of Newton's 2nd Law (here $F=F_o/\gamma$) predicts the same canceling proper-forces on our test charge as well.  However if we now move the test charge to the right at a speed $v$, the 2nd Law predicts that the proper-force $F_{o+}$ exerted on our test charge by the stationary {\em positive charges} remains in the direction of their separation, but {\em increases} in magnitude by a factor of $\gamma$.  By symmetry, the co-moving frame-variant force due to negative charges (which now see the test charge as stopped) will have its previous contribution to the proper-force $F_{o-}$ {\em decreased} by a factor of $1/\gamma$.  

This proper-force provides a frame-invariant platform for combining the two frame-variant but otherwise purely {\em electrostatic} forces, allowing us to conclude that the net proper-force experienced by our moving test charge equals the proper-force from the positive charges on the test charge before it began moving ($kqQ/r^2$) times the non-zero difference between $\gamma$ and $1/\gamma$.  Finally  the net frame-variant force on our moving test particle, reduced from the net proper-force again by that factor of $\gamma$, becomes...
\begin{equation}
F_{up} = \frac{1}{\gamma} (\gamma - \frac{1}{\gamma}) k \frac{q Q}{r^2} = k \frac{q Q}{r^2} \frac{v^2}{c^2} = q v \left( \frac{k}{c^2} \frac{I ds}{r^2} \right)
\end{equation}

This force, due to a frame-dependence of forces in spacetime that has nothing to do with electrostatic forces {\em per se}, is of course traditionally explained by saying that the current creates a magnetic field $B$ according to the Biot-Savart prescription in parentheses on the right, which in turn exerts a force according to the Lorentz Law $F = qvB$.  Thus magnetic fields are a convenient tool for taking into account relativistic effects of the Coulomb force (most noticable around neutral current-carrying wires), and  
the flat-space (one-map two-clock) version of Newton's 2nd law provides us with a derivation that (except for the symmetry invocation) requires only one map-frame with yardsticks and synchronized clocks.

\bibliography{anyspeed.bib}

\end{document}